# Atomic scale investigation of grain boundary structure role on deformation and crack growth dynamics in Aluminum

I. Adlakha[1], M.A. Bhatia[1], K.N. Solanki[1*] and M.A. Tschopp[2]

[1] School of Energy, Matter and Transport; Arizona State University; Tempe, AZ 85287, USA
[2] Weapons and Materials Research Directorate, Aberdeen Proving Ground, MD 21005, USA
*(480)965-1869; (480)727-9321 (fax), E-mail: kiran.solanki@asu.edu, (Corresponding author)

**Abstract**

The role that grain boundary (GB) structure plays on the plasticity of interfaces with preexisting cracks and on the interface crack dynamics was investigated using molecular dynamics for both <100> and <110> aluminum symmetric tilt grain boundaries (STGBs). In simulations with a crack at the interface, this research shows how the maximum normal strength of the interface correlates with the respective GB energy, the GB misorientation, and the GB structural description. For instance, the normal interface strength for GBs containing 'D' structural unit (SU) or stacking faults in the GB structural description ($\Sigma 13$ (510) $\theta=22.6°$ and $\Sigma 97$ (940) $\theta=47.9°$) shows a noticeable decrease in interface strength, as compared to other evaluated <100> GBs that contained favored SUs. In the case of <110> interfaces, the presence of the 'E' SU in the GB structure lowers the maximum normal interface strength by ~35%. Further investigation of the deformation at the crack tip in GBs containing the 'E' structure revealed that the 'E' SU underwent atomic shuffling to accommodate intrinsic stacking faults (ISFs) along the interface, which in turn acts as a site for partial dislocation nucleation. Interestingly, regardless of GB misorientation, GB interfaces examined here containing the 'E' structure in their structural period exhibited relatively small variation in maximum normal strength of interface. The GB volume ahead of the crack tip underwent structural rearrangement which, in turn, influenced the crack propagation mechanism. In most GBs, the crack propagation was due to alternating mechanisms of dislocation emission, followed by propagation of dislocation (blunting) and cleavage/crack advance. Moreover, the crack growth rates along the GB interface were strongly influenced by the initial free volume at the interface, i.e., faster crack growth was observed along interfaces with higher initial free volume. A strong asymmetry in crack growth due to differences in available slip planes ahead of both the crack tips (e.g., $\Sigma 13$ (510), $\Sigma 97$ (940), $\Sigma 11$ (113) and $\Sigma 27$ (552) STGBs) was observed with the crack growth process, with the $\Sigma 97$ (940) $\theta=47.9°$ STGB displaying the largest asymmetry. In most cases, dislocation emission along various available slip planes ahead of the crack tip in <100> STGBs was the dominant plastic event, whereas the <110> STGBs had a partial dislocation emission {111}<112> followed by twin formation for interfaces containing the 'E' SU. Last, there was no direct correlation between any one individual interface constitutive parameters (e.g., free volume, availability of easy slip systems, GB energy, etc.) and interface crack dynamics. Nevertheless, simulation results did indicate an indirect role of these constitutive parameters on atomistic deformation ahead of the crack tip, which significantly alters the interface strength and properties. In summary, the present simulations provide new insight into crack growth along interfaces and provide a physical basis for determining the incipient role between the GB character and interface properties as an input to higher scale models.

# 1. Introduction

Recently, several studies using atomistic methods have dedicated considerable effort towards understanding the brittle versus ductile behavior of materials, including the dynamic instabilities that occur during fracture (Yip and Wolf, 1991; Farkas et al., 2001; Hai and Tadmor, 2003; Li and Chandra, 2003; T. Zhu et al., 2004; Y. T. Zhu et al., 2004; Zhang and Paulino, 2005; Buehler and Gao, 2006; Yamakov et al., 2007; Yamakov and Glaessgen, 2007; Bobylev et al., 2010; Solanki et al., 2011; Adlakha et al., 2013; Han and Tomar, 2013; Lee and Tomar, 2013). In particular, some of these studies focus on the deformation process at the crack tip, i.e., brittle (cleavage) versus ductile (nucleation of partial *and* full dislocations, and deformation twinning) behavior and dislocation burst in single crystal and bicrystal metals. These simulations have clarified the effects of applied load orientation and slip planes on dislocation nucleation or twinning at/ahead of the crack tip, both in the single crystal and at the grain boundaries (GBs). In general, atomistic simulations have helped in understanding that crack tip propagation is primarily governed by the dislocation process (emission versus cleavage) and in some metals, such as aluminum and copper, deformation may be accommodated through deformation twinning (Liao et al., 2003; Warner et al., 2007; Yamakov and Glaessgen, 2007; Xiong et al., 2012). Deformation twinning, especially in aluminum, has been a subject of debate due to discrepancies in experimental and atomistic modeling results, regardless of the time and length scale issues (Farkas et al., 2001; Hai and Tadmor, 2003; Warner et al., 2007; Yamakov and Glaessgen, 2007). For example, Tadmor and Hai (Tadmor and Hai, 2003) used a quasi-continuum method to suggest that deformation twins form ahead of the crack tip in single crystal aluminum when $T = \lambda_{crit}\sqrt{\gamma_{us}/\gamma_{ut}} > 1$, where $\lambda_{crit}$ is the normalized critical load for the nucleation of a trailing partial dislocation, $\gamma_{us}$ is the energy associated with unstable stacking fault energy, and $\gamma_{ut}$ is the unstable twinning partial nucleation energy. Using atomistic simulations, Farkas et al. and others (Farkas et al., 2001; Yamakov et al., 2002; Y. T. Zhu et al., 2004; Li et al., 2009) observed similar twinning behavior in Al, while it has been well-established experimentally that Al does not twin except under certain loading conditions and at relative small time scales, primarily due to the high stacking fault energy of the material (Pond and Garcia-Garcia, 1981; Warner et al., 2007; Li et al., 2009). On the other hand, this discrepancy does not occur in nanocrystalline (NC) metals (grain size < ~100 nm) (Yamakov et al., 2002, 2003a; Chen et al., 2003; Wolf et al., 2005; Liao et al., 2003; Li et al., 2009), where deformation twinning is indeed one of the experimentally-observed deformation mechanisms, along with the GB rotation/sliding and dislocation glide (Chen et al., 2003; Liao et al., 2003; Li et al., 2009). This behavior may be partially attributed to GB interfaces and networks that contribute significantly to the properties of metals. In fact, Yamakov et al. (Yamakov et al., 2002, 2003b; Wolf et al., 2005) show roughening of GBs due to the dislocation-GB interaction and the formation of a new grain due to twin emission/absorption/reformation. Furthermore, onset of plastic deformation is further complicated due to the dissociated and faceted structure present in GB interfaces, and general Schmid *and* non-Schmid factors alone do not adequately account for such plasticity events (Spearot et al., 2007; Tschopp and McDowell, 2008). Hence, the local structure and character of the GB was important for understanding interfacial crack mechanics.

Research has shown that the physical properties of GBs are affected by both the macroscopic degrees of freedom and the microscopic local structure associated with GBs (Sutton and Vitek, 1983a; Wang et al., 1984; Gui-Jin and Vitek, 1986; Rittner and Seidman, 1996). In terms of macroscopic degrees of freedom, the term GB character is often used. The GB character refers to the five degrees of freedom associated with the misorientation between the crystallographic orientations of the two adjoining grains. Hence, GB character encompasses not only the misorientation angle, but also the GB plane as well as information pertaining to GB type (e.g., low angle versus high angle, Σ value, etc.). In terms of microscopic local structure, the translations between adjoining grains are important, as is the localized dislocation structure of the boundary. Historically, many efforts focus on developing a method to characterize GBs (Burgers, 1940; Bragg and Nye, 1947; Read and Shockley, 1950; Bishop and Chalmers, 1968; Hirth, 1974; Grimmer et al., 1974) and their influence on the physical properties of polycrystalline material. These models utilized dislocation arrays, disclinations, and coincident site lattice (CSL) to describe microscopic and macroscopic degrees of freedom of GBs. Based on identifying the favored GB for the corresponding GB systems, these methodologies (Bristowe and Crocker, 1978; Pond, 1979; Pond et al., 1979; Sutton and Vitek, 1983a; Ingle and Crocker, 1980; Vitek et al., 1980; Sutton, 1982, 1989; Rittner and Seidman, 1996) described the structural elements comprising symmetric tilt, asymmetric tilt, twist, and twin GBs. They determined that the favored GBs are entirely composed of unique structural units (SUs) that cannot be decomposed into other GB structures.

There have been several studies that examine the role of GB structural units on GB plasticity and deformation mechanisms. For instance, Sansoz and Molinari (Sansoz and Molinari, 2005) used quasi-continuum simulations on Al and Cu bicrystals to show that deformation sliding was accommodated through atomic rearrangement of 'E' SUs in the GB structure for the <110> symmetric tilt grain boundary (STGB) system. Additionally, Spearot et al. (Spearot et al., 2005, 2007) have used molecular dynamics simulations at 10 K and 300 K to examine how GB structural units impacted the process of dislocation nucleation. However, the role of GB character on plastic events such as dislocation emission, twin formation, and atomic shuffling at the interface with a preexisting crack has received less attention, especially in Al with varying degrees of GB SUs. Furthermore, there is no clear structure-property relationship that exists between the crack dynamics and the GB character. In fact, experimentally in Cu and Fe bicrystals it has been shown that there is a strong directional dependence on the fracture behavior along the interface (Wang and Anderson, 1991; Saeedvafa, 1992; Wang and Mesarovic, 1995; Kysar, 2000, 2001; Tang et al., 2005) and the growth asymmetry is governed by the relative angle of the slip plane to the crack plane and the growth direction. However, complete understanding of the fundamental mechanisms associated with asymmetric deformation are inaccessible with experimental techniques, especially the atomic deformation ahead of the advancing crack and resulting asymmetric crack growth behavior. Thus, this presents significant challenges to developing unique state variables (e.g., free volume, GB energy, SU present along the grain boundary etc.) capable of being hierarchically transferred to higher-order models for predictive modeling.

Therefore, the objective of the present research was to understand the atomistic relationship between the local structure and the crack mechanics at the GB interface. In this work, molecular dynamics (MD) simulations of Al bicrystals at 300 K were used for various <100> and <110> tilt grain boundaries to clarify the role of interface character on dislocation emission, formation of twins, and/or atomic shuffling. Of particular interest was how the GB character affects the crack dynamics (emission of dislocation, twin or rearrangement, and asymmetry in the crack growth) and the instability in the crack motion. The GB was modeled with a center crack along the GB interface, and it was subjected to uniaxial tensile loading normal to the GB interface, as depicted in Figure 1a. The SUs present in the GB were correlated with the observed fracture behavior. The crack growth during the fracture process was calculated and a notable variation in growth was observed for the +X direction and –X direction (from the center of the specimen).

This paper is organized as follows. Our simulation approach is briefly summarized in Section 2. The results and discussion section describe the atomic structure of GBs at 0 K, the stress-strain responses of the interfaces, the crack growth along both +X and – X directions of each interface, and the atomic deformation for GB interfaces displaying interesting crack growth features. The simulation results reveal several interesting observations: (1) there was a strong asymmetry in crack growth due to difference in available slip planes ahead of both the crack tips (in particular, the $\Sigma 13$ (510), $\Sigma 97$ (940), $\Sigma 11$ (113), and $\Sigma 27$ (552) STGBs), (2) in some cases, the GB volume ahead of the crack tip underwent structural rearrangement, which subsequently influences the crack propagation mechanism; (3) in most GBs, crack propagation was caused by alternating mechanisms of dislocation emission, followed by propagation of dislocation (blunting) and cleavage/crack advance, (4) the crack growth rates along the GB interface were strongly influenced by the initial free volume at the interface, i.e., faster crack growth was observed along interfaces with higher initial free volume, (5) the 'E' SU in <110> STGBs shows a distinct deformation behavior compared to other <110> GB interfaces, primarily due to the ease of nucleating the intrinsic stacking fault (ISF) ahead of the crack tip, which acts further as a site for nucleating a partial dislocation, and (6) during analysis of the <110> STGBs, we observed that twinning occurs only in the GB interfaces with 'E' SU in the GB structural description. In summary, this work provides a framework for analyzing GB deformation when primarily influenced by the SUs. This approach can be used to inform non-local cohesive zone models of atomistic grain boundary details in an effort to capture the relevant interfacial physics for higher-length scale models.

## 2. Computational Methods

A parallel molecular dynamics code (large-scale atomic/molecular massively parallel simulator, LAMMPS (Plimpton, 1995)) with a semi-empirical embedded atom method (EAM) (Daw and Baskes, 1984) potential was used to study the role of GB character on plastic events and the subsequent crack dynamic behavior in Al STGBs with <100> and <110> tilt axes. In this work, we employed the EAM potential of Liu et al. (Liu et al., 2004) to describe the interactions between Al-Al and for generating the impurity-free GB systems. The EAM potential was fitted to the experimental data and a set of *ab-initio* structural atomic forces calculated for large set of configurations. Furthermore, this potential has been shown to accurately reproduce the experimental solid-liquid binary phase diagram (Buehler, 2008), thus quantifying stable structure formations and thermo-mechanical behavior of nanocrystalline materials (Wolf et al., 2005) and elucidating the mechanistic causes for the crossover from a dislocation to a grain boundary-based deformation mechanism with decreasing grain size, grain boundary structure and energies, and many other physical behaviors, including the stacking fault energy (SFE) in Al (Buehler, 2008). The SFE is a critical intrinsic material property when modeling extended defects such as dislocations and GBs.

The GB structure and minimum energy were calculated using a bicrystal simulation cell with three-dimensional (3D) periodic boundary conditions consisting of two grains at 0 K as described by Rittner and Seidman (Rittner and Seidman, 1996). The periodic boundaries were maintained with a separation distance of 12 nm between the boundaries. The several 0 K minimum energy GB structures were obtained through successive rigid body translations followed by an atom deletion technique and energy minimization using a non-linear conjugate method (Rhodes et al., 2013; Rittner and Seidman, 1996; Solanki et al., 2013; Tschopp and McDowell, 2008; Tucker and McDowell, 2011).

The minimized GB structures with geometries as illustrated in Figure 1a were modeled by prescribing a periodic condition along the Y and Z directions and a non-periodic condition along the X direction. Prescribing a periodic boundary condition along the Y direction introduced a second GB interface at the end of the simulation setup, but the influence of this interface on the results was minimized by prescribing sufficient separation between the two GB planes. The dimensions along the X and Z directions were chosen to provide sufficient atomic layers for convergence of the 3D interfacial fracture. The atomistic model was equilibrated at 300 K using a canonical ensemble (NVT) for 5 ns and the periodic directions were subsequently relaxed using the isothermal-isobaric (NPT) equations of motions for 10 ns. Then, a through-thickness crack was introduced along the GB plane as illustrated in Figure 1a, such that the 2$a$/$W$ ratio for the atomistic model was kept constant at 0.1, where 2$a$ is the crack length, and $W$ is the total width of the specimen (X dimension, ~50 nm). The neighboring interactions among atoms of the top and bottom halves of the crack were not included during deformation to avoid spurious interactions of atoms across the crack width. The minimum dimensions for the entire bicrystal at equilibrium were approximately 50 nm × 50 nm × 5 nm (~1.0 million atoms). The model was deformed under tensile loading by applying a constant strain rate of in $10^8$ s$^{-1}$ perpendicular to the GB plane (Y direction), while the transverse boundaries were controlled using the NVT ensemble equations of motion. Each bicrystal interface was deformed to an engineering strain of 8%.

Continuum measures, such as stresses and strains, were calculated using the virial definition (Tsai, 1979). Note that virial stresses can be significantly different from continuum stresses at the free surfaces. The evolution of the defected atoms (e.g., the crack tip, dislocations, and twins) was visualized using the centrosymmetry parameter (Kelchner et al., 1998). This method provides accurate identification of defects present and the ability to separate them from regions of large elastic deformation, which retain the symmetry of their local environment during elastic deformation. The crack tip propagation was tracked using the centrosymmetry parameter; a cutoff of the centrosymmetric parameter was used to identify surface atoms (~20) and track the crack trajectory. Hence, crack growth could be separately tracked in the +X and –X directions.

Atomic rearrangement that occurs during minimization of the GB can be expressed as relative displacement perpendicular to the GB plane (Chen et al., 1989) (Figure 1b). This provides a microscopic measure for the excess volume or the free volume required to create the interface because the strain field associated with the interface at the atomistic scale decays as $e^{-z}$, where z is the distance normal to the GB plane. The free volume can be measured by finding the relative displacement of two parallel atomic planes nearest to the GB. This measure represents the local expansion undergone by crystals at the interface ($\Delta z^*$). This measure is relatively difficult to quantify using traditional experimental approaches due to limitations in atomic resolution. Hence, experiments measure the net expansion of the two grains, i.e., by measuring the relative displacement of atomic planes over a larger distance from the interface. The GB free volumes were analyzed using this method. This free volume measure gives an ability to correlate crack dynamics along interfaces with scalar variables such as the free volume.

## 3. Results and Discussion

The minimum energy GB structural description was characterized using the structural unit model (SUM) (Smith et al., 1977; Pond et al., 1979; Sutton and Vitek, 1983a, 1983b; Vitek et al., 1983), and the initial free volume of each interface was calculated using the previously-described method. The tensile stress-strain response of the <100> and <110> STGBs with a middle crack was analyzed to correlate the mechanical response to the GB SUs. Subsequently, plastic events and crack growth dynamics were evaluated along the + X and –X directions of the GB interface. Finally, the implication of atomic structure on the GB deformation (i.e., slip versus cleavage behavior) was analyzed for various GB interfaces.

### 3.1. Characteristics of GBs: Structural unit, GB energy, and free volume

Understanding the structure and energy of the GB system is crucial for engineering materials intended for advanced applications because GB properties can vary widely (coherent twin versus low angle versus high angle GBs). In this study, a range of GB structures and energies that are representative of some of the variation observed in the GB character distribution of polycrystalline as well as nanocrystalline metals was used to investigate the role of GB character on plastic events such as dislocation emission, twin formation, and atomic shuffling at the interface with a preexisting crack.

### 3.1.1 <100> STGB

Five <100> STGB configurations with various misorientation angles were created in Al at 0 K. The GB energy and SU period were calculated for these configurations at 0 K along with the initial free volume of the GBs (Table 1). Table 1 also lists the maximum interface normal strength, which will be subsequently discussed in Section 3.2. The calculated GB energies were comparable to what has been previously reported in the literature (Hasson et al., 1970; Wang et al., 1984; Sansoz and Molinari, 2005; Tschopp and McDowell, 2007). The structure-energy correlation can provide more details pertaining to the variation in GB energies, as each GB has characteristic SUs that describe its atomistic morphology. Low-angle boundaries can be represented by an array of discrete dislocations spaced a certain distance apart. However, at higher misorientation angles (high-angle GBs), the dislocation spacing is small enough that dislocation cores overlap and dislocations rearrange to minimize the boundary energy. The resulting GB structures are often characterized by GB dislocations or structural units (Sutton and Vitek, 1983a) as shown in Figure 2. Here, the interface structures were characterized by viewing them along the <100> tilt axis. In Figure 2, atoms are colored based on their position along the tilt axis (Z direction). Grain boundaries with certain misorientation angles (and typically a low $\Sigma$ value) corresponded to "favored" boundaries, while other boundaries were characterized by SUs from the two neighboring favored boundaries.

The GB structures in the <100> STGB system are shown in Figure 2. The two $\Sigma 5$ boundaries can be described, in the sense of Sutton and Vitek (Sutton and Vitek, 1983a), as "favored" STGBs because each GB has characteristic structural units describing its atomistic morphology. The $\Sigma 13$ (510) $\theta=22.6°$ GB (Figure 2a) is composed of one 'C' unit and two 'D' units along the periodic length of the GB. The SU for the $\Sigma 5$ (310) $\theta=36.9°$ STGB is labeled as 'C' and it consists of 9 atoms and has a SU periodic length of 12.80 Å, as shown in Figure 2b. The Σ97 (940) θ=47.9° GB structure is more complex, as shown in Figure 2c, in that it has a distorted variation of the SU as reported by Gui-Jin and Vitek (Gui-Jin and Vitek, 1986). The B'' structural element is elongated along the X direction. The C' has an atom missing, unlike the favored 'C' SU, and it is elongated along the X direction. The Σ97 (940) STGB has the highest free volume and SU period length: $0.251a_0$ and 39.08 Å, respectively. The Σ5 (210) θ=53.1° GB structure, as shown in Figure 2d, is composed of a "B'" SU with a SU periodic length of 9.05 Å. The 'dot' in the notation signifies that the SU shifted from {001} to the {002} neighboring plane. These types of structures were classified as 'centered' by Sutton (Sutton, 1989), or sometimes even referred to as mirror symmetry. The energy and the free volume of the Σ5 (210) STGB are 565 mJ/m$^2$ and $0.0945a_0$, respectively, and for the Σ5 (310) STGB, 551 mJ/m$^2$ and $0.167a_0$, respectively. Figure 2e shows the centered structural unit of Σ13 (320) θ=22.6° GB, which is composed of |AB'.AB'|, where the A and B' consist of 4 atoms arranged in a diamond and have a SU period length of 14.60 Å. The energy and the free volume of two Σ13 GBs are as follows: for the Σ13 (510) STGB, 542 mJ/m$^2$ and $0.1754a_0$, and for the Σ13 (320) STGB, 481 mJ/m$^2$ and $0.225a_0$, respectively. Table 1 presents a summary of the investigated <100> GBs with corresponding GB energies and the initial free volumes of the interface. These structures and energies are in agreement with previous studies on low CSL symmetric and asymmetric tilt grain boundaries in Al and Cu (Tschopp and McDowell, 2007a, 2007b).

### 3.1.2 <110> STGB

The SUs associated with each investigated <110> STGB are shown in Figure 3 and summarized in Table 2, along with the 0 K GB energy, the free volume, and the SU periodic length of each interface. As with Table 1, Table 2 also lists the maximum interface normal strength detailed in Section 3.2. The GB SU descriptions are in good agreement with previous descriptions available in literature (Rittner and Seidman, 1996; Sansoz and Molinari, 2005; Spearot et al., 2007; Tschopp et al., 2007, 2008a) . The Σ3 (111) coherent twin boundary with the 'D' SU and the Σ11 (113) STGB with the 'C' SU form the major cusps in the energy-misorientation angle for the <110> tilt axis. The 'A' and 'C' SUs consist of 4 atoms each connected in a diamond, and the 'D' SU is a Shockley partial dislocation that is found at the end of a SF connecting it to the interface for lower misorientation angles and is the primary SU for the Σ3 (111) STGB. The 'E' SU is made up of 6 atoms arranged in a kite shape. The presence of the 'E' SU creates a larger free volume when compared to other SUs for the <110> tilt axis, which can result in very different grain boundary structures and free volumes as shown in (Tschopp et al., 2007).

The preceding SU characterization was important for determining the (strong) influence of the GB SUs on the maximum normal stress. Specifically, the maximum normal interface strength during tensile loading correlates with the GB energy as a function of the misorientation angle, more so for the <100> system (with an exception of the Σ97 (940)) than for the <110> system. Also, the presence of 'E' SU was associated with a particular type of deformation behavior at the interface, as will be explained in the following sections.

### 3.2. Mechanical response of various GB interfaces

This section details MD simulations of GBs with a center crack used to compute the stress-strain response normal to the GB interface for various GBs under tension.

### 3.2.1 <100> STGB

Figure 4a shows the average tensile response of <100> GB interfaces as a function of applied normal strains and corresponding maximum stresses are tabulated in Table 1. The simulation results indicate the SU influence on the mechanical response for the GB interfaces. For example, the stress decrease following to the maximum normal interface strength was larger for the Σ97 (940) STGB compared to other GB interfaces. This larger decrease in stress can be attributed to a low resistance to crack growth due to the higher free volume (hence, less plasticity). Additionally, as tabulated in Table 1, the maximum normal interface strength for GBs with 'D' or SF present, such as Σ13 (510) and the distorted SU present in Σ97 (940) were lower, about 18.8% and 25.2%, respectively, when compared with the Σ5 (210) STGB. It should be pointed out that the Σ97 (940) has the highest GB energy of all the <100> STGBs analyzed here, about 24% higher than the Σ13 (320) STGB. Figure 4b shows various normalized constitutive parameters, such as GB energy and maximum normal interface strength as a function of the misorientation of the <100> STGBs under tension. The GB energy and the maximum normal interface strength were normalized with 596 mJ/m$^2$ (Σ97 GB) and 2.86 GPa (Σ5 (210) GB), respectively. Qualitatively, the maximum normal interface strength during tensile loading correlates directly with the GB energy, except for the Σ97 (940) STGB, which has the highest initial free volume, periodic length, GB energy (Table 1), and the lowest maximum normal interface strength when compared with the other <100> STGBs examined here. The latter observation (lowest maximum normal interface strength) is not surprising, given that the Σ97 (940) STGB structure is more complex, highly distorted, *and* consists of a combination of B" and C' SUs.

### 3.2.2 <110> STGB

The mechanical response under tension of <110> STGBs as a function of applied normal strains is shown in Figure 5a and corresponding maximum stresses are tabulated in Table 2. The trend shown in Figure 5a is different from the <100> STGBs trend depicted in Figure 4a. In the case of <110> STGBs, the mechanical behavior can be grouped into two distinct sets, i.e., one for GBs with a structural period containing 'E' SU and another one for GBs without 'E' SU. This type of distinct behavior leads to the fact that the presence of 'E' SU lowers the dislocation nucleation stress. The maximum interface strength for all of the <110> STGBs evaluated was between 1.62 GPa and 2.70 GPa (Σ9 (221) and Σ11 (113), respectively), suggesting a strong effect of the GB structure. Qualitatively, the *average* maximum normal interface strength for GBs containing the 'E' SU, as tabulated in Table 2, is about 1.63 GPa as compared to other non-'E' SU GBs which is about 2.52 GPa.

The GBs containing the 'E' SU show a lower stress response than GBs with other SUs. Figure 5b shows various normalized constitutive parameters, such as GB energy and maximum normal stress as a function of misorientation of <110> STGBs under tension. The GB energy and the maximum normal interface strength were normalized with 474 mJ/m$^2$ and 2.7 GPa, respectively. Our results indicate no particular trend for the <110> STGBs under tension when comparing various interface constitutive parameters, i.e., maximum normal interface strength, GB energy, etc., except presence of 'E' SU causes a large drop, on average 35%, in the maximum normal interface strength for <110> STGBs (see Figure 5b and Table 2).

### 3.3. Directional crack growth dynamics and the influence of GB structural units

This section describes how the interfacial fracture behavior of <100> and <110> STGBs was analyzed to characterize the effect of the GB SU and crack dynamics, on the crack growth behavior.

**3.3.1 <100> STGB**

Figures 6a and 6b show the directional crack growth behavior along the interface from both the left (-X direction) and right (+X direction) sides of the middle tension crack, respectively, for different <100> STGB systems. Table 3 quantifies two important figures of merit for crack growth of different <100> STGBs: the maximum crack growth velocity and the corresponding time for rapid growth. The crack growth behavior of <100> STGBs, as shown in Figure 6, shows a significant crack growth asymmetry between the left and right sides of the initial crack with stronger influences of relative orientation between the slip plane to the crack plane, the growth direction, and interface SUs. Furthermore, during the deformation, some interface cracks had a longer inactive (blunting) interval, followed by varying continuous growth with discrete jumps and final failure, which was indicative of the alternating mechanisms of slip and cleavage. For example, as shown in Figure 6, the $\Sigma 5$ (210) GB exhibited an initial crack growth rate of 57 m/s (t = 375-410 ps) and 50 m/s (t = 380-450 ps) for the left and the right sides, respectively. During this event, the crack growth front advanced through interfacial separation, as will be further explained later by analyzing the intrinsic deformation mechanisms (Section 3.4). Subsequent to this initial rapid growth event, the crack growth retarded to a rate of 28 m/s (t =410-430 ps) along the -X direction and then to negligible rates beyond 450 ps along the +X direction due to slip activity. Similarly, for the $\Sigma 5$ (310) GB, the initial crack growth rate is 27 m/s (t = 340–400 ps) and 50 m/s (t = 380-400 ps) along the -X and +X directions, respectively. After which, the crack tip blunts due to dislocation activities and negligible crack growth was observed along both the –X and +X directions. For all the <100> STGBs evaluated here, the slowest crack growth was observed along the –X and +X directions (8.5 m/s and 27 m/s) for the $\Sigma 13$ (320) GB.

The GB containing 'D' SU or SF in their structural period description, such as $\Sigma 13$ (510) and $\Sigma 97$ (940), exhibited different crack dynamics compared to other GB interfaces. As shown in Figures 6a and 6b, the $\Sigma 13$ (510) had the initial crack growth rates of 30 m/s (t = 290-375 ps) and 50 m/s (t = 290-390 ps) along the –X and +X directions, respectively. During these periods, the crack propagation was accompanied by atomic shuffling (shown in Section 3.4). Figures 6a and 6b show the growth rate for the second region from 425 ps to 460 ps was about 100 m/s along the –X direction and 55 m/s along the +X direction (450–540 ps). Similar behavior was also observed for the $\Sigma 97$ (940) GB, where the interface had higher initial free volume compared to the $\Sigma 13$ (510) GB. The crack growth along the +X direction (~75 m/s) was comparatively higher with regards to the growth along the –X direction, where the growth was halted from 350-400 ps due to a dislocation nucleation. The Σ97 (940) GB interface failed in a brittle manner along the +X direction around 425 ps (ε = 4.25%).

These distinctions in computed crack growth rates at various stages of the GB fracture process were attributed to several factors: the slip plane orientation relative to the crack plane and the growth direction, the initial GB free volume, and the GB SUs. For instance, the asymmetry of the GB SUs themselves (shown in Figures 2 and 3) may manifest as different initial resistances to crack growth depending on the initial crack direction. For the Σ5 (210), Σ5 (310), and Σ13 (320) GBs, the growth asymmetry diminished after adequate time elapsed in the plastic region. However, the growth activity for the Σ13 (510) varied on each side, as marked by the slow crack growth regions for both the left and right crack tips (375-425 ps and 400-450 ps), as shown in Figures 6a and 6b. The large variation in crack growth could be attributed to the relative orientation of slip system ahead of the crack tips for these interfaces as well as to the free volume of the GB.

### 3.3.2 <110> STGB

In the case of <110> STGB interfaces, the crack growth along both directions was negligible for many interfaces that contained 'E' SU in their structural period description (see Figure 7). Most of the interfaces with 'E' SU in their structural period description experienced the following sequence of deformation behavior at both crack tips, which resulted in negligible crack growth (blunting). There are several stages of deformation observed in the <110> STGBs under tension. First, with an increase in the applied strain, atomic shuffling along the GB interface led to formation of an ISF, followed by nucleation of a leading partial dislocation. The leading partial was soon followed by a twin partial dislocation along the adjacent slip plane to form the twin embryo, and then further application of strain lead to twin growth. An interesting feature was noted, especially in the Σ33 (441) GB, in that the crack growth along the +X and –X directions was negative for long period during the deformation process, implying that the crack healed. This behavior is attributed to the nucleation of partial dislocations, which subsequently resulted in deformation twins. The analysis of the atomic mechanisms responsible for this deformation will be further discussed in Section 3.4.

The crack growth asymmetry was evident in the Σ33 (225) interface as shown in Figure 7. For example, following an inactive period for up to 270 ps, the stable crack growth rate along the –X direction was about 30 m/s when compared to ~20 m/s for the +X direction. Moreover, the crack growth along the –X direction was arrested from 420-575 ps due to plastic events ahead of the crack tip. In contrast, the growth in the Σ33 (225) along the +X direction did not experience a prolonged growth arrest (530-570 ps) when compared to the growth behavior along the –X direction. The Σ33 (225) GB contains a high initial free volume across the interface, which was credited for the rapid growth rate (~80 m/s) in the presence of inaccessible slip systems in front of the crack tip, i.e., cleavage type behavior.

The Σ27 (552) had a large asymmetry in the directional crack growth behavior, especially after 300 ps, as shown in Figure 7. This type of crack behavior was due to heterogeneous deformation at both sides of the crack tip. For example, along the –X direction, the nucleation of partial dislocation was observed around 160 ps, which subsequently lead to the twin nucleation. On the other hand, the partial dislocation nucleated around 200 ps along the +X direction. The crack growth along the +X direction was in burst after a certain period of arrest (400-500 ps and ~550-600 ps), and atomic shuffling was the predominant mode for crack growth during these bursts. The asymmetry in the crack growth based on these results is significantly affected by available easy slip planes, the initial free volume, and the interface SU. Nevertheless, the 'E' SU had a significant role on the interface crack dynamics and properties when compared to other SUs, and this behavior was explored further in the next section by examining atomic deformation mechanisms along the GB interface. Table 4 provides highlights of maximum growth periods experienced by each interface and the maximum crack tip velocity along both +X and –X directions in the case of <110> STGB interfaces. Interestingly, these simulations quantify that not only do different grain boundary structures and structural units influence crack growth, but different crack growth directions can also heavily influence crack growth rates as well. This asymmetry in crack growth rates within a single grain boundary and its correlation with the atomic structure of the interface has not been as well documented in the literature.

### 3.4. Analyzing deformation mechanisms to explain mechanical response and crack dynamics

A more detailed analysis of crack dynamics and asymmetry in crack growth was performed by visualizing the atomic region along the interface ahead of the moving crack tip. The centrosymmetry parameter was used to detect dislocations and stacking faults, which have larger centrosymmetry values than atoms within a pristine crystal. The four prominent deformation modes observed in the following study are highlighted below.

### 3.4.1 Dislocation emission, low free volume, <100> STGB

The centrosymmetry (csym) of atoms around the crack tip along the +X direction of the Σ13 (510) GB with its time evolution is shown in Figures 8a-g. At low applied strain, atoms in front of the crack tip along the interface retained a large part of their initial SU description (see Figures 8a-c). However, as the applied strain increased, a clear crack growth behavior was observed, accompanied with a large amount of atomic shuffling (see Figures 8d-f). Most of the atomic rearrangement observed here was localized to the GB interface and its vicinity. The steady state crack growth was observed in the absence of dislocation nucleation along the +X direction up to an applied strain of $\varepsilon = 3.6\%$ (Figure 8f), during which atomic shuffling was the dominant mechanism. Subsequently, a partial dislocation was nucleated along the +X direction at the applied strain of $\varepsilon = 3.7\%$ (Figure 8g) along the $(\bar{1}\bar{1}1)[\bar{2}1\bar{1}]$, which resulted in crack arrest (see Figure 6b). In contrast, the dislocation nucleation event for the crack tip along the –X direction occurred at an earlier applied strain of $\varepsilon = 3.25\%$, as shown in Figure 8i. Also notable regarding the deformation along both directions was the varying degree of atomic shuffling that was required for each. The +X direction underwent a large amount of atomic shuffling prior to dislocation nucleation, in contrast to minimal shuffling involved with the deformation along –X direction.

### 3.4.2 Dislocation emission, high free volume, <100> STGB

Next, the deformation evolution of Σ97 (940) GB was examined as shown in Figure 9. This interface had the highest initial free volume, periodic length, GB energy, and the lowest maximum normal interface strength when compared with other <100> STGBs examined here. The steady state crack growth was observed with a crack velocity of 75 m/s (Figure 6b) in the absence of slip activities along the +X direction as the tensile loading was applied (see Figures 9a-d). The majority of work done by the applied stress along the +X direction of the crack tip was transferred to decohesion of the interface. Meanwhile, dislocation nucleation was observed on the $(\bar{1}\bar{1}1)[112]$ slip system ahead of the crack tip along the –X direction (Figure 9h). The crack growth along the –X direction was inhibited by dislocation emission. The asymmetric crack growth behavior as observed in Figures 9d & h was largely due to accessibility of slip systems ahead of the crack tip. Similar results were reported by Yamakov et al. (Yamakov et al., 2007) for Σ99 GB in Al. The crack growth rate showed an increase along the highly distorted SUs of the Σ97 (940) GB, which also had the highest initial free volume.

### 3.4.3 Partial dislocation emission, <110> STGB

In Sections 3.2 and 3.3, we examined the mechanical responses and crack dynamics of various <110> interfaces under uniform tension, where we observed interfaces with the 'E' SU exhibited behavior different from other boundaries. The atomic deformation mechanisms associated with the 'E' SU were analyzed to understand the behavior of crack growth along both +X and –X directions for these <110> STGB interfaces. Most interfaces identified in Figure 7 (Σ33 (441), Σ9 (221) and Σ11 (332)) showed a large amount of crack tip plasticity. Dislocation nucleation ahead of the crack tip occurred at low stresses along both interface directions to accommodate deformation. In most of the <110> GBs analyzed here, the atomistic deformations had the following characteristics: either (i) nucleation of the ISF or a larger amount of atomic shuffling followed by a nucleation of partial dislocation, or (ii) in some cases, twinning before interface failure was witnessed (especially for GBs with 'E' SU). The discrete nature of non-homogenous atomistic deformation observed here was the predominate cause behind the growth asymmetry. For example, the crack growth along both directions of the crack tip in the Σ11 (113) interface was fairly asymmetric (see Figures 7 and 10). On closer analysis (Figure 10), the incipient strain of the plastic event along each crack tip varied greatly. At the applied strain of 2.6% (Figure 10c), an ISF was nucleated ahead of the crack tip along the +X direction; subsequently, a partial dislocation $(\bar{1}\bar{1}1)[\bar{1}\bar{1}\bar{2}]$ was nucleated from the ISF (Figure 10d). Also, the partial dislocation nucleation in the top grain was followed by partial dislocation nucleation along $(111)[11\bar{2}]$ in the bottom grain, as shown in Figure 10f. An important proponent to deformation was the small realignment of the 'C' SU in creating the ISF, as shown in Figures 10(a-d).

On the other hand, along the –X direction, a steady state crack growth was observed up to an applied strain of 3.5% (Figures 7a and 10g), followed by nucleation of an ISF. The growth was arrested along the –X direction between applied strains of 3.5-4.6%, primarily due to blunting of the crack tip and due to the relative angle between the initial slip system and the crack plane requiring a large amount of energy to propagate the dislocation away from the crack tip. Subsequently, the crack tip along the –X direction (Figure 10i) witnessed partial dislocation nucleation in both the crystals. A large amount of atomic shuffling was witnessed at the interface to nucleate partial dislocation along the –X direction (inset Figure 10h), in contrast to the minimal atomic movement that was required for the slip nucleation along the +X direction. This was a key aspect of the atomic deformation of the Σ11 (113) GB interface, and the principal cause for the crack asymmetry observed during the failure of the interface as shown in Figure 7a.

### 3.4.4 Deformation twinning, <110> STGB

The atomic deformation behavior of the Σ27 (552) GB interface was analyzed to help understand deformation twinning in the presence of 'E' SU in the <110> GB interfaces, as shown in Figure 11. Along the +X direction, a slight rearrangement of the 'E' SU in the interface leads to nucleation of an ISF, as shown in Figure 11c. With an increase in the applied strain to about 2%, nucleation of the leading partial dislocation along $(111)[11\bar{2}]$ from the ISF was observed, as shown in Figure 11d. Dislocation emission occurred at a much lower applied strain ($\varepsilon = 2\%$) than in other <110> GB interfaces. The extent of atomic shuffling required to nucleate the partial dislocation was negligible. The leading partial was trailed by a twin partial dislocation to form the twin embryo (see Figure 11e). As the applied strain increased, a growth of a twin was observed as shown in Figure 11f.

The atomic deformation ahead of the crack tip along the –X direction is shown in Figure 11g-k. Unlike the deformation behavior along the +X direction, significant atomic shuffling along the interface was observed to accommodate the initial nucleation of the ISF (see Figure 11g-i). The nucleation of $(111)[11\bar{2}]$ partial dislocation at $\varepsilon = 1.7\%$ left behind a SF in its wake, as shown in Figure 11j. This was followed by an emission of the twin partial and growth of a twin (Figure 11k). The analysis of the deformation behavior ahead of the crack tip along both +X and –X directions shows that relatively minimal applied strain was required to cause the 'E' SU to rearrange and form an ISF along the interface. This ISF acted both as an accelerant in the plastic events at the interface and as a source for nucleating partial dislocations, which were followed by deformation twins. Also, these discrete nature of plastic events were significantly affected by the combination of availability of easy slip planes, the free volume, and SU - notably not just the SU or the GB energy alone (Tschopp et al., 2008b).

To further address the role of the 'E' SU, the key events during the deformation of Σ27 (552) GB on the left (-X direction) and right (+X direction) sides of the middle tension crack were correlated with the mechanical response of the interface, as shown in Figure 12. This figure reveals some key trends which can be described as follows. As the applied normal strain increased 1) the nucleation of twin partial dislocation occurred along the –X direction of the crack as indicated in step 1 ($\varepsilon = 1.8\%$); 2) then the nucleation of twin partial dislocation along the +X direction occurred as depicted in step 2 ($\varepsilon = 2.3\%$); 3) during segments 2-3, twin growth occurred; 4) subsequently at step 3 ($\varepsilon = 2.9\%$), another twin dislocation nucleated along +X direction; 5) during segments 3-4, deformation twins grew; 6) after step 4 ($\varepsilon = 3.4\%$), the twin partial dislocation was nucleated along the –X direction ahead of the crack tip; and 7) finally, at step 5 ($\varepsilon = 3.7\%$ ), nucleation of partial dislocations occurred along both –X and +X directions.

### 3.4.5 Comparison of deformation modes in <100> and <110> STGBs

Overall, the 'E' SU in <110> GB exhibits unique deformation behavior (see Table 5), and we observed direct correlation between the presence of the 'E' SU in the structural period and maximum normal interface strength (see Figure 5b). Predominantly, the <110> GBs with 'E' SUs in their structural period exhibited deformation through atomic shuffling of the interface atoms generating an ISF, which resulted in a partial dislocation nucleation from the generated ISF, followed by twinning. Not all <110> interfaces examined in the present study developed a deformation twin and, typically, <110> interfaces that formed a deformation twin showed a lower maximum strength when compared to interfaces with no twin. On the other hand, all

<100> GB interfaces examined here underwent deformation by dislocation emission along the available slip planes ahead of the crack tip (Table 6). The <100> GB interfaces displayed appreciably higher crack growth rates (Table 3), in contrast to <110> GBs interfaces (Table 4). Therefore, the GB local arrangements and resulting structural units, such as the 'E' SU present in the GB structural period, alter the deformation around/ahead of the crack tip, the crack growth along the interface is influenced by the accessibility to slip systems ahead of the crack tip, and the rate of crack propagation was influenced by the initial free volume of the GB.

## IV. Conclusion

In this work, we used MD to investigate the role of GB character on plastic events, such as dislocation emission, twin formation, and atomic shuffling, at the interface with a preexisting crack and the interface crack dynamics in <100> and <110> STGBs. Of particular interest was how the GB character affects crack dynamics and propagation. Specifically, we noted the amount that the crack grows will depend upon the directions within the GB that it propagates, and the local atomic scale deformation will alter the overall growth rate and behavior. This research shows that the collective contribution from the GB SUs, the availability of easy slip planes, and the initial free volume plays a significant role in the interfacial deformation and crack dynamics (see Tables 1-4). The maximum interface normal strength for the <100> GB examined here directly correlates with the respective GB energy. In the case of the <110> system, we observed two distinct responses ($\theta < 109.48°$ and $\theta > 109.48$) of interface maximum normal strength and the GB misorientation.  We observed no direct correlation between individual interface constitutive parameters, such as free volume, availability of easy slip systems, GB energy, etc., and interface crack dynamics. However, results presented here indicate an indirect role of these constitutive parameters on the atomistic deformations ahead of the crack tip, which significantly alter the interface strength *and* properties. Furthermore, no unique parameter can be used to describe/model the interface behavior. The following conclusions from this work can be summarized as follows:

1. The normal interface strength for GBs containing 'D' SU or SF in the GB structural description ($\Sigma 13$ (510) $\theta=22.6°$ and $\Sigma 97$ (940) $\theta=47.9°$) showed noticeably lower interface strength compared to other <100> GBs evaluated that contained favored SUs in the GB structural description (see Figure 4).
2. The stress-strain response for the <110> STGB interface (Figure 5) shows that the presence of the 'E' SU lowers the maximum normal interface strength by an average 35% when compared to other <110> STGBs. Upon further investigation of the deformation at the crack tip in GBs containing the 'E' structure, the 'E' SU underwent atomic shuffling to accommodate ISF along the interface and ISF acts as a site for the partial dislocation nucleation. The increased presence of 'E' in the GB structure decreased the GB's maximum interface strength by nearly 5% when compared to other 'E' SU GBs (see Figure 5a, Table 2). Furthermore, regardless of GB misorientation, GB interfaces examined here that contained 'E' structures in their structural period exhibited relatively similar maximum interface strength (see Figure 5b, Table 2).
3. The $\Sigma 13$ (510) $\theta=22.6°$ and $\Sigma 97$ (940) $\theta=47.9°$ GBs exhibit an asymmetric directional crack growth process, while the rest of the <100> GBs show relatively symmetric growth behavior. The directional crack growth process for a $\Sigma 97$ (940) $\theta=47.9°$ interface displayed the greatest asymmetry (Figures 5 and 9). The maximum growth rate witnessed by the GB interface was in correlation with the initial free volume of the GB interface (Tables 1 and 5), implying that GBs with higher free volume experienced greater growth rates with cleavage being the dominant mechanism rather than dislocation slip. However, in most GBs, the crack propagation was due to alternating mechanisms of dislocation emission followed by propagation of dislocation (blunting) and cleavage/crack advance.
4. The majority of <110> STGB interfaces shows negligible crack growth along both the +X and –X directions (Figure 6 and Table 4). On the other hand, a few GB interfaces (Σ27 (552), Σ11 (113), and Σ33 (225)) shows a large amount of asymmetric crack growth behavior due to the relative difference in angle between the slip plane to the crack plane and the growth direction. The growth rate increased with the initial free volume of interface (Tables 2 and 6).
5. The nucleation, propagation and interaction of partial dislocations with other dislocations ahead of the crack tip in <100> STGBs were the major plastic events (Table 6), whereas the <110> STGBs had a partial dislocation emission {111} <112> followed by a twin formation for interfaces containing the 'E' SU (Table 5).
6. The directional crack growth dynamics along the interface was governed by orientation of slip planes relative to the crack plane and the growth direction, the initial free volume, and the interface SU. Moreover, we observed that combinations of these constitutive parameters have a larger role than any individual parameter does alone. Nevertheless, the 'E' SU has a significant role on the interface crack dynamics and deformation behavior when compared with other SUs. For example, during analysis of the <110> STGBs, we observed the unanticipated and intriguing process that twinning occurred only in the GB interfaces that contained 'E' SUs in GB structural description (Figure 11, Table 5). In fact, experimentally (Wang and Anderson, 1991; Wang and Mesarovic, 1995; Kysar, 2000) it has been shown that the directional crack growth response in certain GBs could vary greatly from ductile to brittle depending on the availability of slip systems ahead of the crack tip.
7. Finally, the GB structural unit plays a large part in dislocation nucleation ahead of the crack tip, as interface atoms provide varying degrees of mobility to incorporate plastic deformations. In particular, the crack growth rate along the interface correlated with the initial free volume of the GB. In summary, these new atomistic perspectives provide a physical basis for recognizing the incipient role between the GB character and interface properties, including GB energies as an input to higher scale models (Idesman et al., 2000; Bammann and Solanki, 2010; Solanki and Bammann, 2010). For example, the present study provides a stepping stone to development of a physics-based, atomistically informed non-local cohesive zone model of interfaces that can accommodate the required atomic properties, including local deformations, to be used in finite element models to enhance predictive capability.

**Acknowledgement**

The authors would like to recognize Dr. W. Mullins and Dr. A.K. Vasudevan from the Office of Naval Research for providing insights and valuable suggestions. This material is based upon work supported by the Office of Naval Research under contract no. N000141110793. We would also like to acknowledge the Fulton High Performance Computing at Arizona State University. MAT would like to acknowledge support from ARL administered by the Oak Ridge Institute for Science and Education through an interagency agreement between the U.S. Department of Energy and ARL.

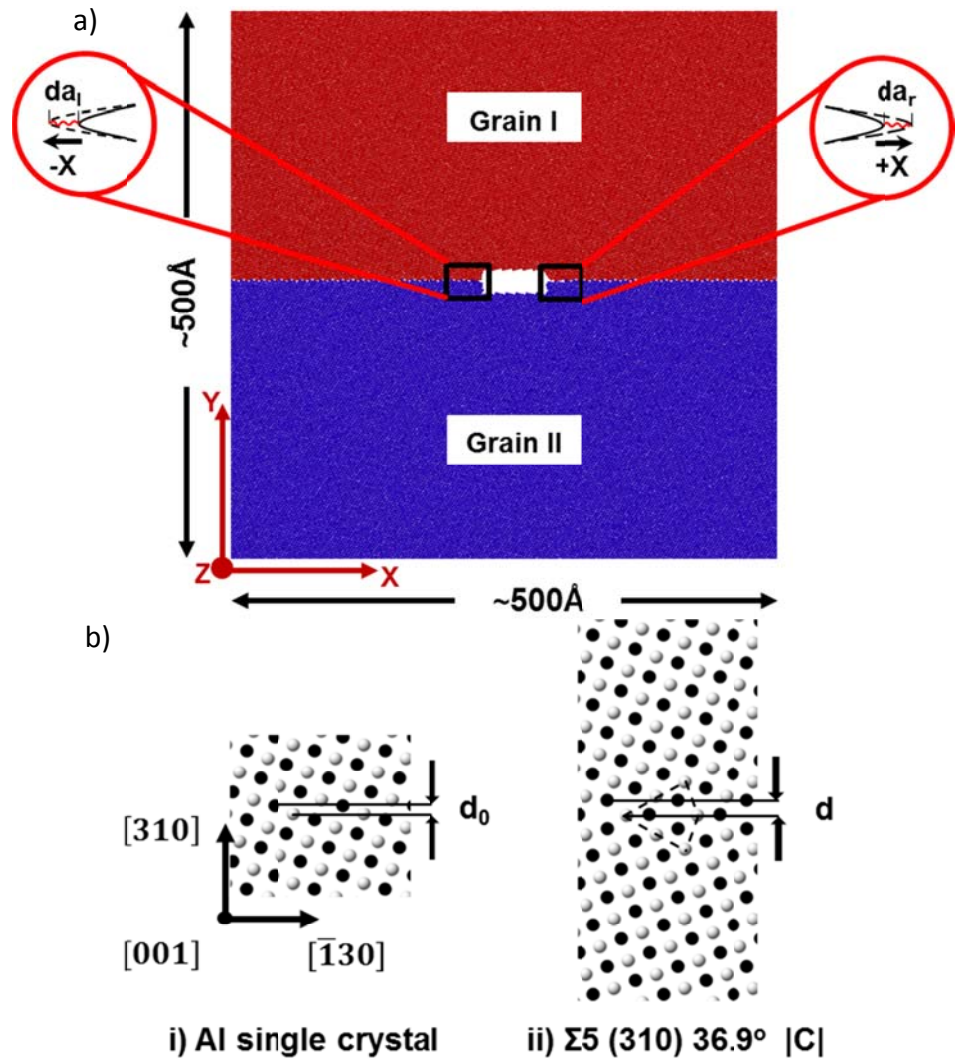

Figure 1. a) Atomistic cell with a GB interface perpendicular to the Y axis. The red and blue regions are grains I and II, respectively. For grains I and II, the Z tilt axis is the same for both, i.e., <100> or <110>. Tensile load was applied along the Y axis to investigate GB SUs' role on plastic events and any subsequent interfacial failure behavior. b) A schematic of GB ($\Delta z^* = d - d_0$) free volume measurement: (i) the measurement of interplanar distance, $d_0$, of the single crystal, and (ii) the interplanar distance, $d$, at the center of the GB plane.

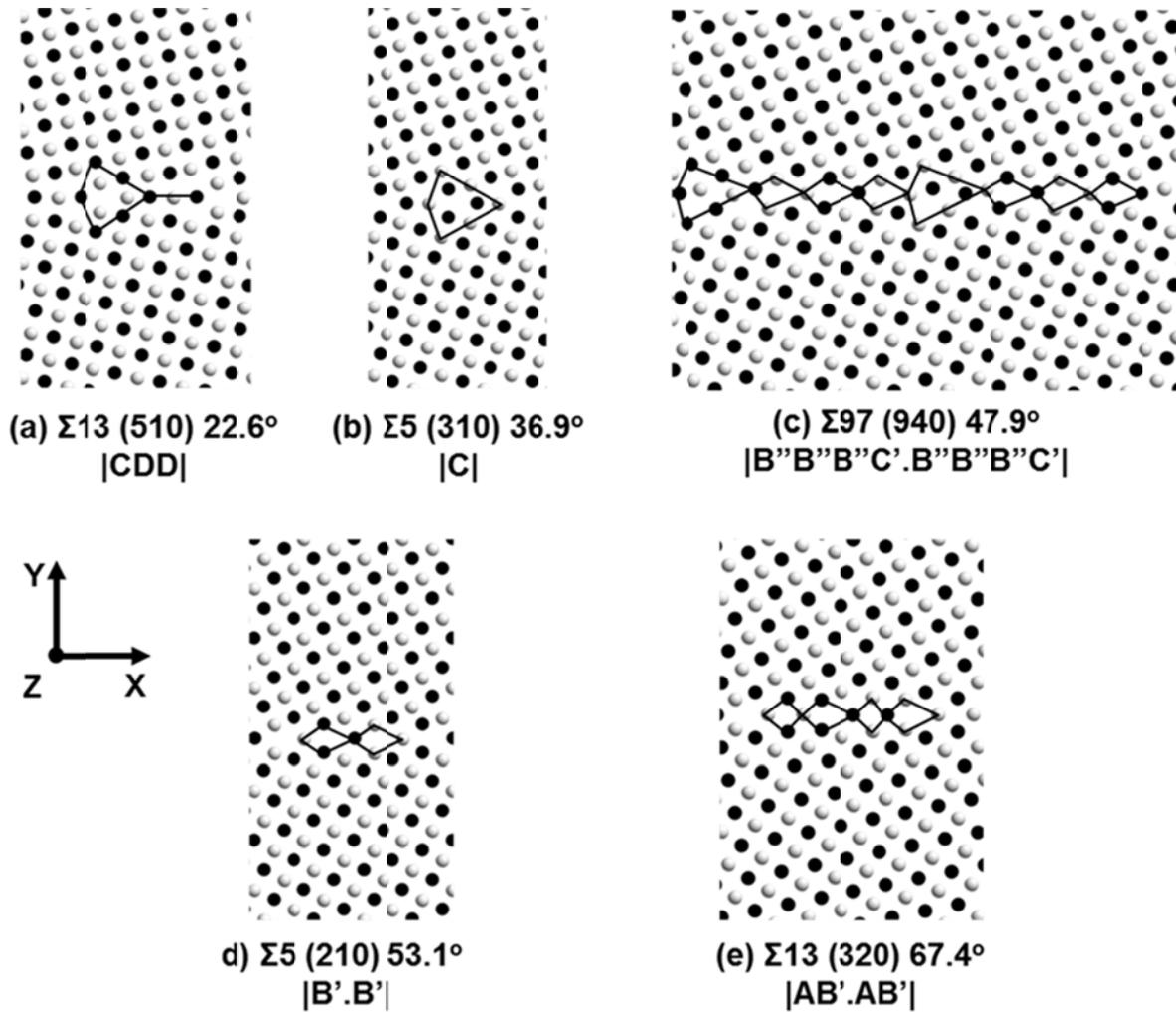

Figure 2. <100> symmetric tilt GB structures with SUs outlined for a) Σ13 (510) θ=22.6°, b) Σ5 (310) θ=36.9°, c) Σ97 (940) θ=47.9°, d) Σ5 (210) θ=53.1° and e) Σ13 (320) θ=67.4°. The white and black circles denote atoms on {001} and {002} planes.

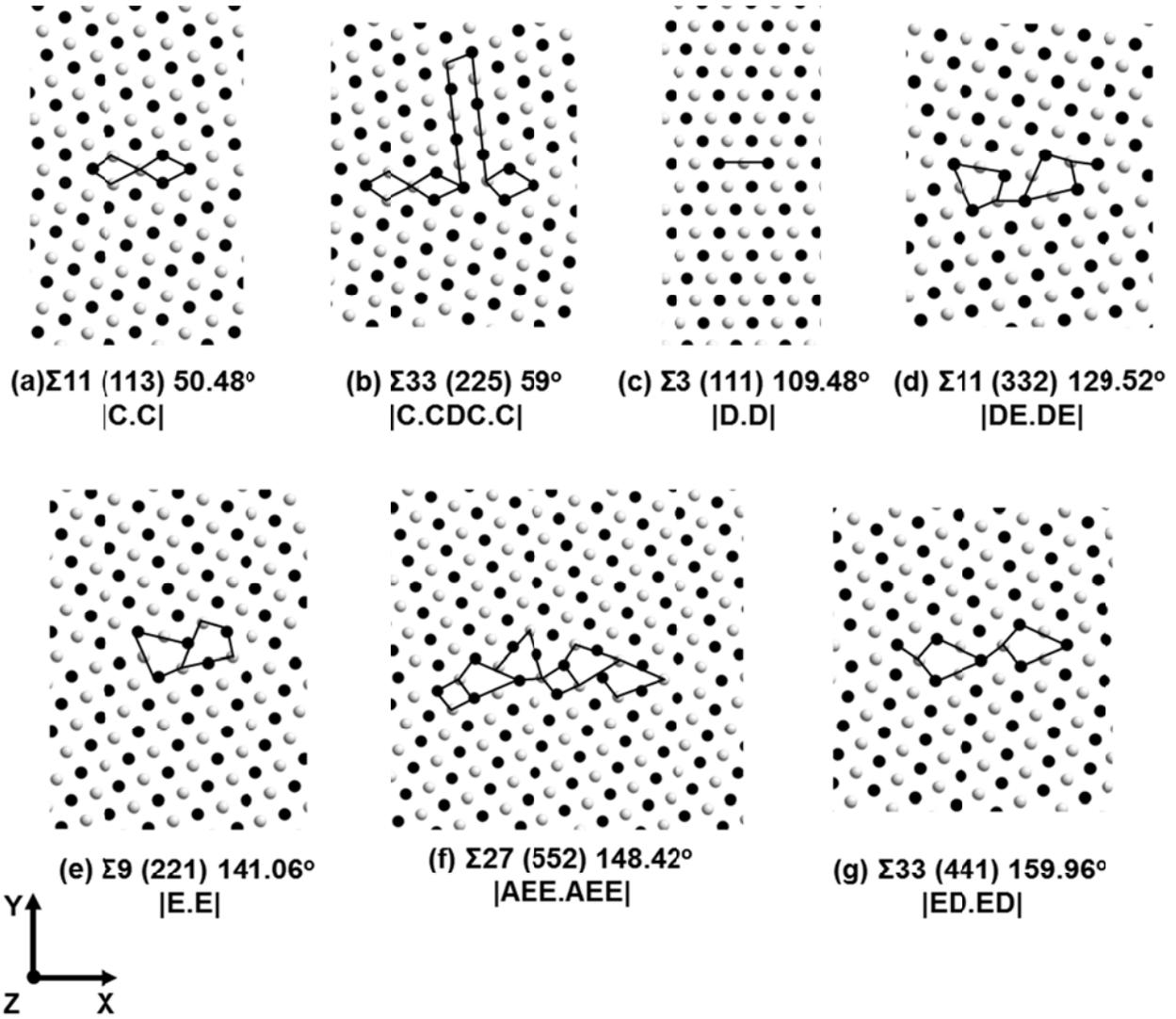

Figure 3. <110> symmetric tilt GB structures with structural units outlined for a) Σ11 (113) (θ=50.48°), b) Σ33 (225) (θ=59°), c) Σ3 (111) (θ=109.48°), d) Σ11 (332) (θ=129.52°), e) Σ9 (221) (θ=141.06°), f) Σ27 (552) (θ=148.52°) and g) Σ33 (441) (θ=159.96°). White and black dots denote atoms on {011} and {022} planes.

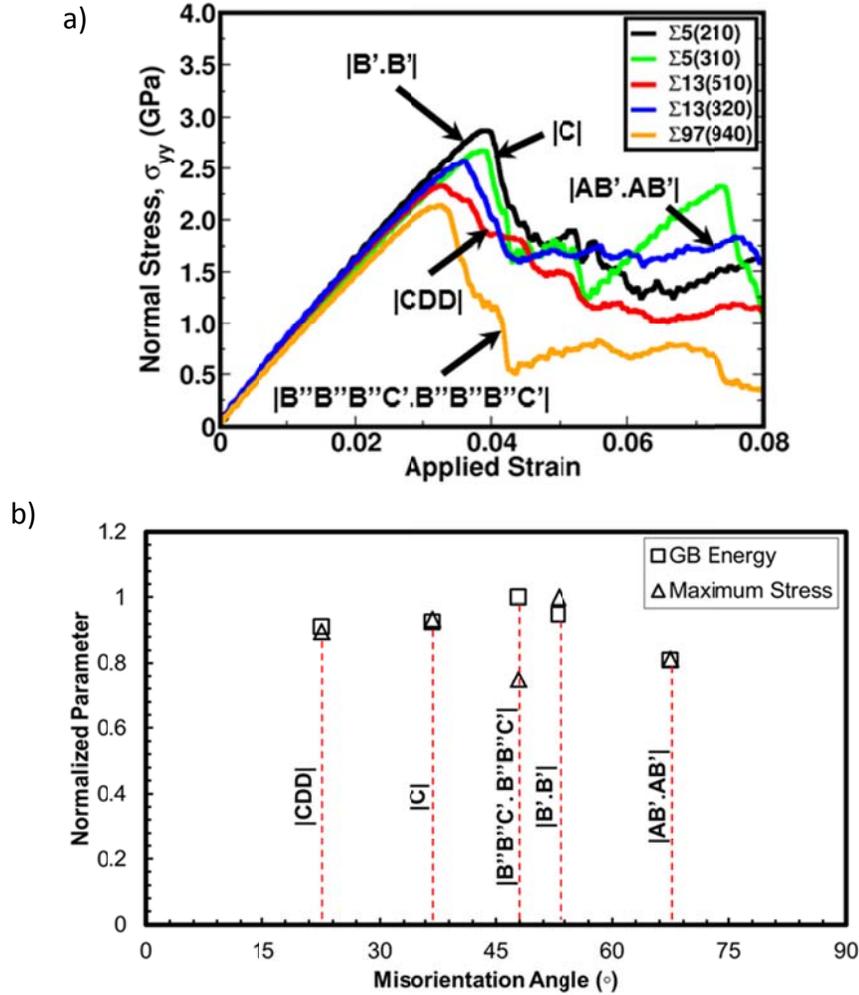

Figure 4. a) Stress-strain behavior of various <100> GB interfaces under uniform tensile loading. The simulation results indicate influence of SU on the maximum normal stress encountered by the GB interface, and b) various normalized constitutive parameters of <100> STGBs under tension. The GB energy and the maximum normal interface strength were normalized with 596 mJ/m$^2$ and 2.86 GPa, respectively.

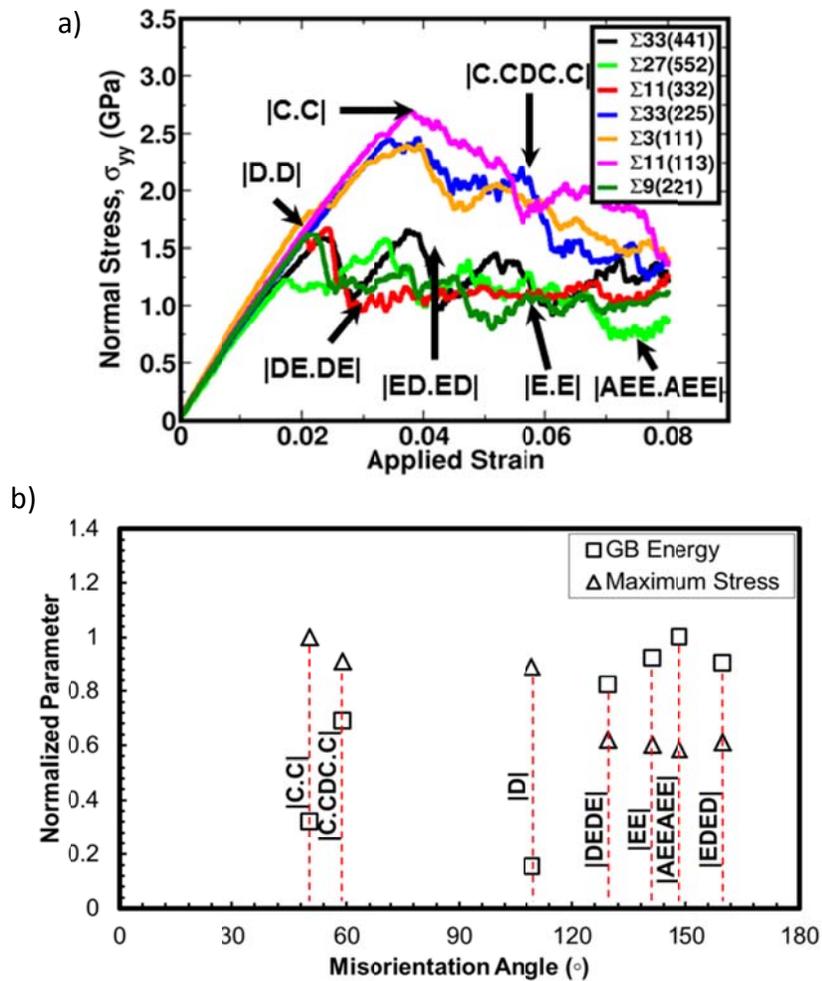

Figure 5. a) Stress-strain behavior of various <110> GB interfaces under tensile loading. The simulation results indicate influence of SU on the maximum normal stress encountered by the GB interface, especially 'E' SU, and b) various normalized constitutive parameters of <110> STGBs under tension. The GB energy and the maximum normal stress were normalized with 474 mJ/m$^2$ and 2.7 GPa, respectively.

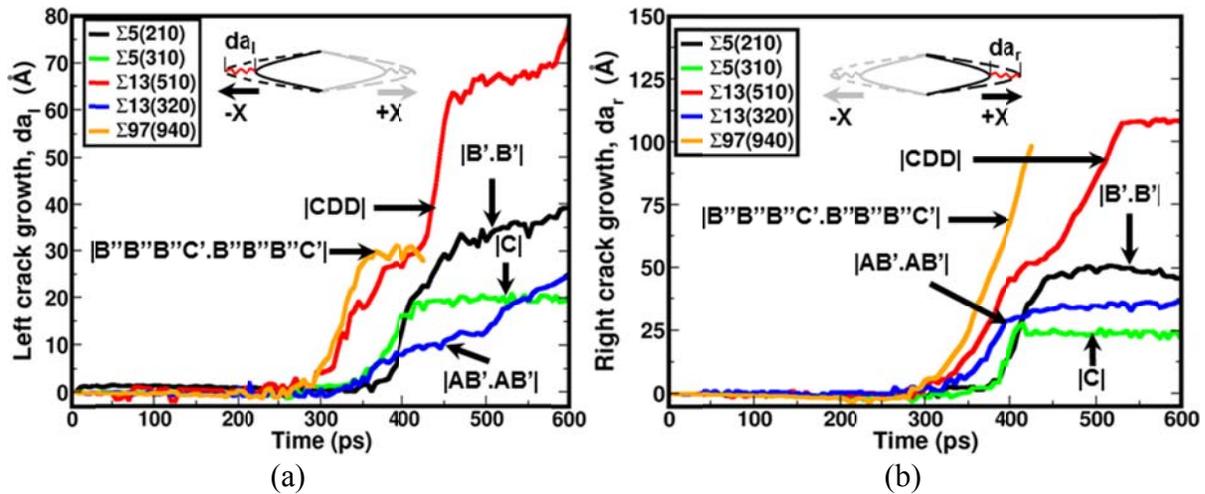

(a)          (b)

Figure 6. The time history evolution of the crack growth in various GBs exhibits distinguishing features, such as inactive, continuous growth, arrest, second continuous growth, followed by final failure: a) the left (-X direction), and b) the right (+X direction) sides of the middle tension specimen.

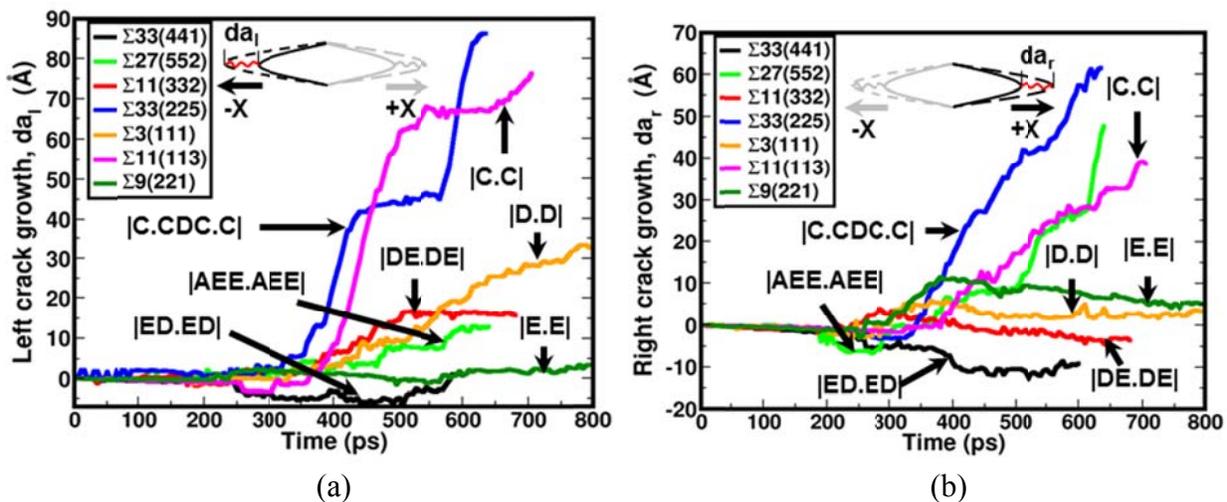

(a)          (b)

Figure 7. The crack growth evolution in various <110> GBs exhibits distinct features, such as inactive, continuous growth, arrest, second continuous growth, followed by final failure: a) the left (-X direction), and b) the right (+X direction) sides of the middle tension specimen.

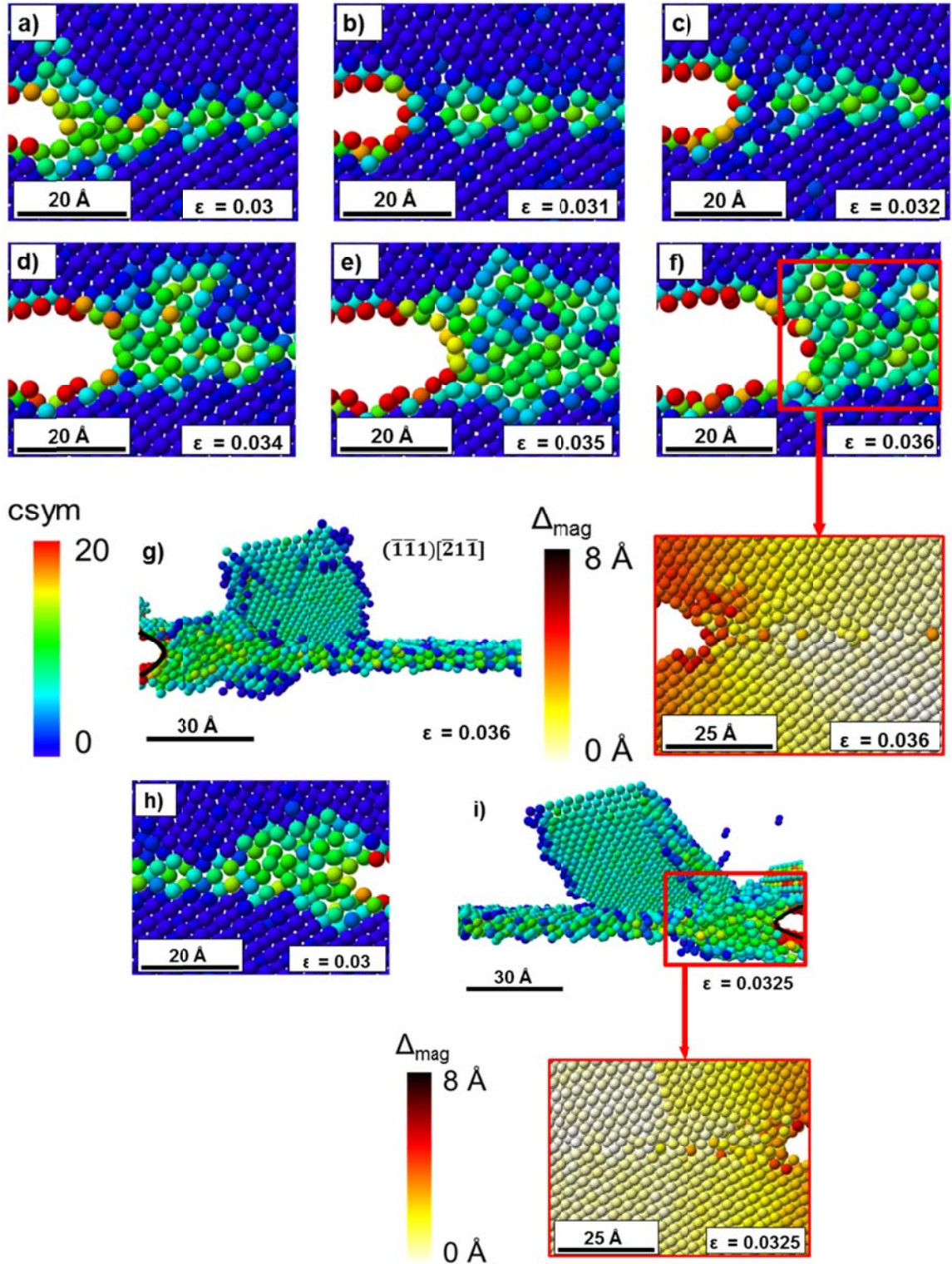

Figure 8. (a-f) Centrosymmetry evolution profile of atoms ahead of the crack tip Σ13 (510) GB along the +X direction; (g) $(\bar{1}\bar{1}1)[\bar{2}1\bar{1}]$ dislocation nucleation ahead of the crack tip along the +X direction (ε = 3.7%) from a 3D perspective (viewing only defected and surface atoms); (h) the crack tip deformation along the –X direction of the Σ13 (510) interface; (i) a partial dislocation nucleation along the –X direction in Σ13 (510) interface (3D perspective). In Figure

8g and 8i, the black curve delineates the crack surface. The inset shows contour of atomic displacement magnitude with respect to the initial configuration, suggesting large atomic movement is required to nucleate a dislocation ahead of the crack tip along both the directions.

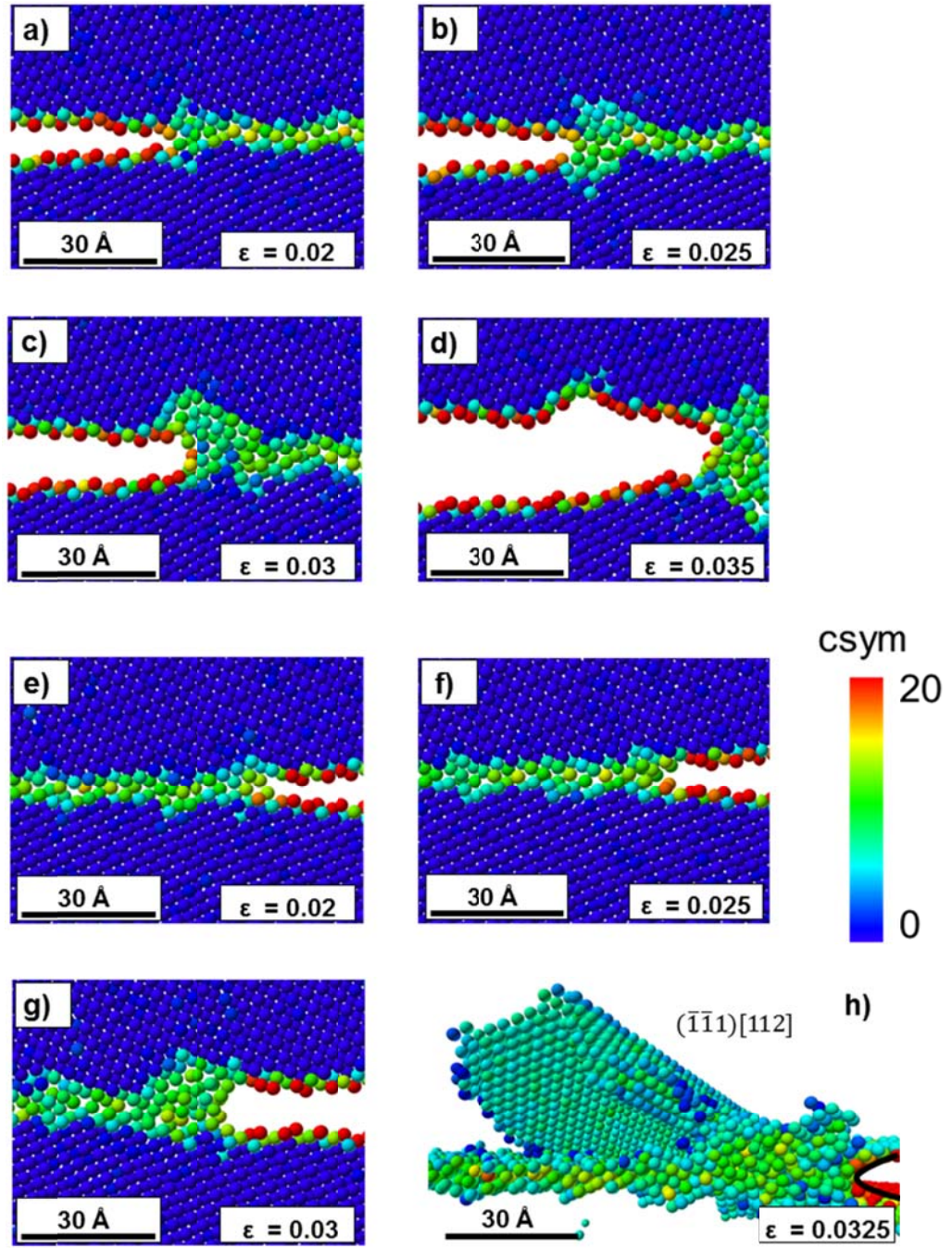

Figure 9. (a-d) Centrosymmetry profile of atoms ahead of the crack tip along the +X direction (right side) of the Σ97 (940) GB, and (e-h) the dislocation nucleation events along the –X direction of the Σ97 (940) GB. The black curve identifies the crack surface.

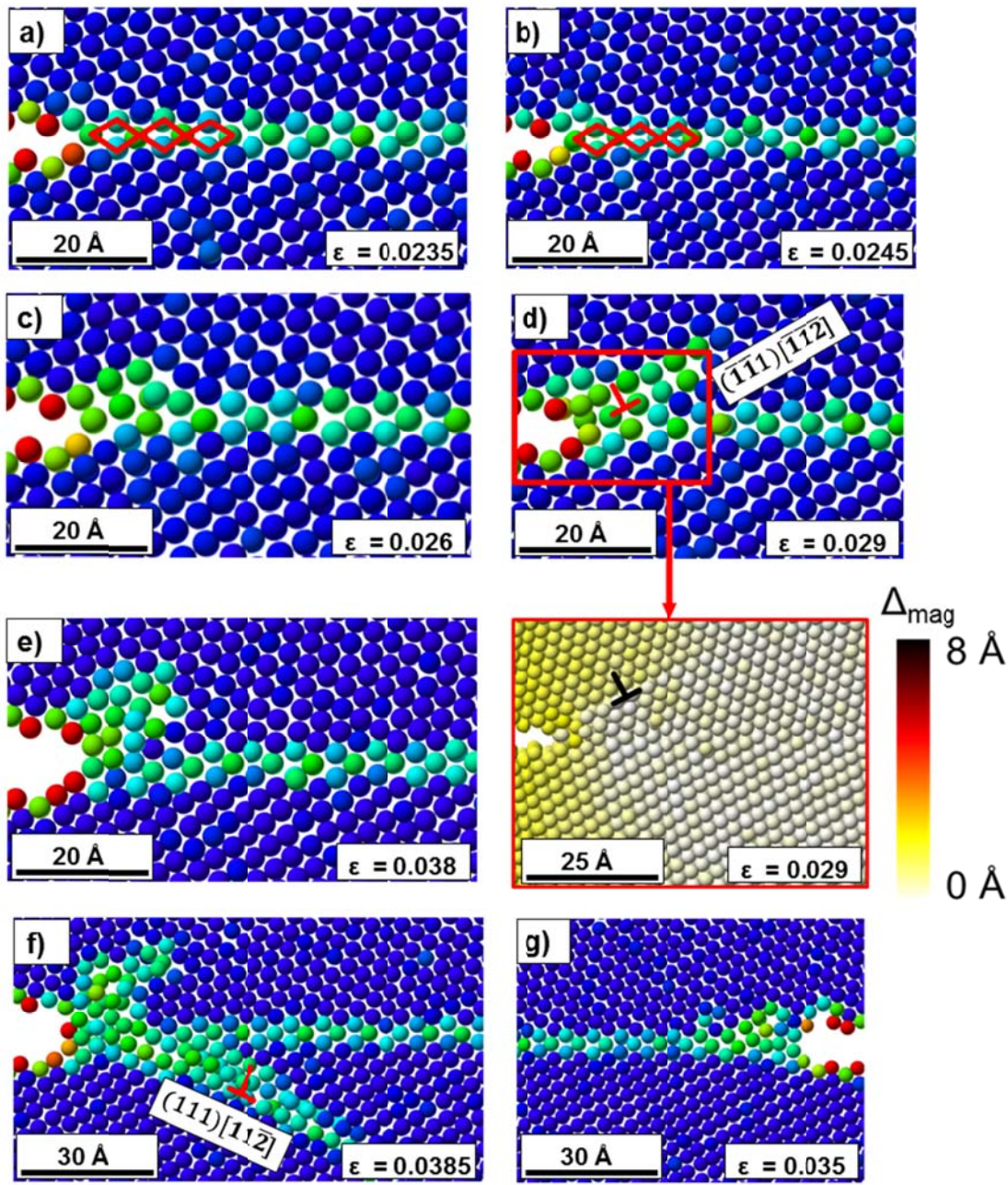

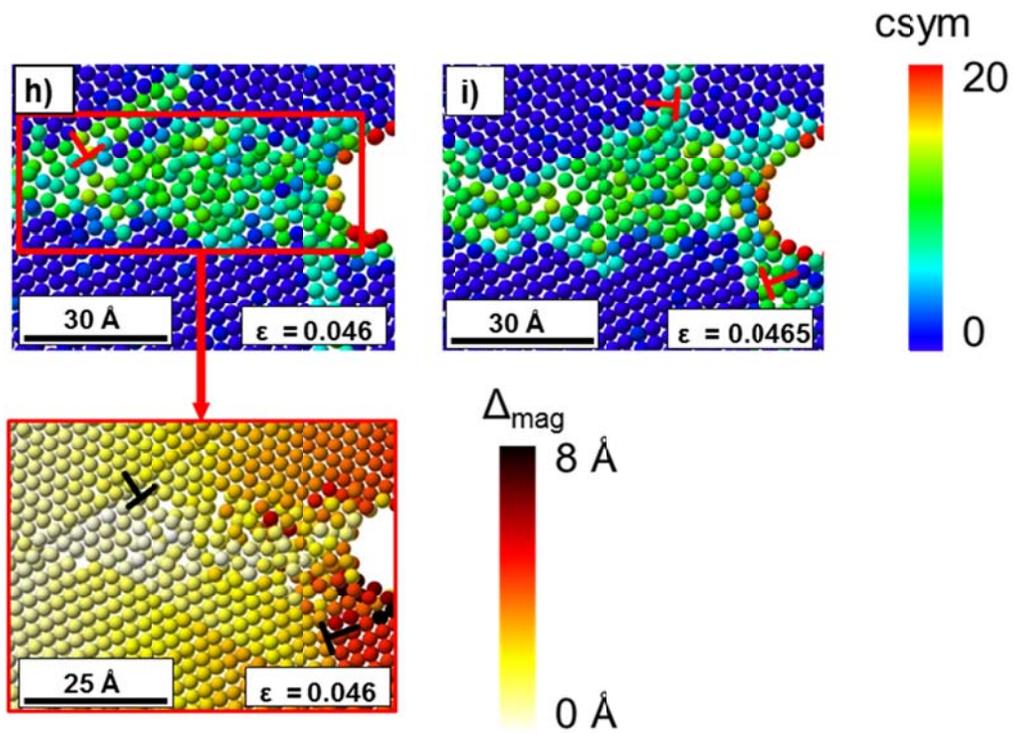

Figure 10. The plastic events across the Σ11 (113) GB interface with atoms were identified using the centrosymmetry parameter. (a-f) show plastic events during the crack growth along the +X direction; (g-i) illustrate the dislocation nucleation during the crack propagation along the –X direction. The inset shows contour of atomic displacement magnitude with respect to the initial configuration, suggesting a large atomic movement is required to nucleate a dislocation ahead of the crack tip along both the directions

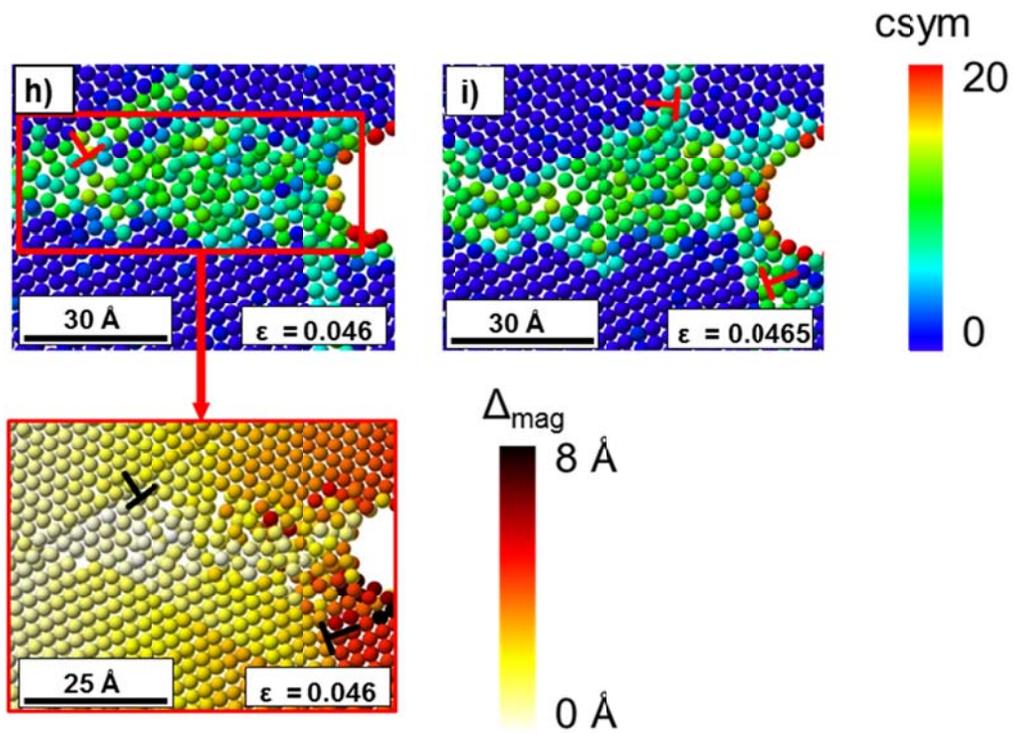

Figure 10. The plastic events across the Σ11 (113) GB interface with atoms were identified using the centrosymmetry parameter. (a-f) show plastic events during the crack growth along the +X direction; (g-i) illustrate the dislocation nucleation during the crack propagation along the –X direction. The inset shows contour of atomic displacement magnitude with respect to the initial configuration, suggesting a large atomic movement is required to nucleate a dislocation ahead of the crack tip along both the directions

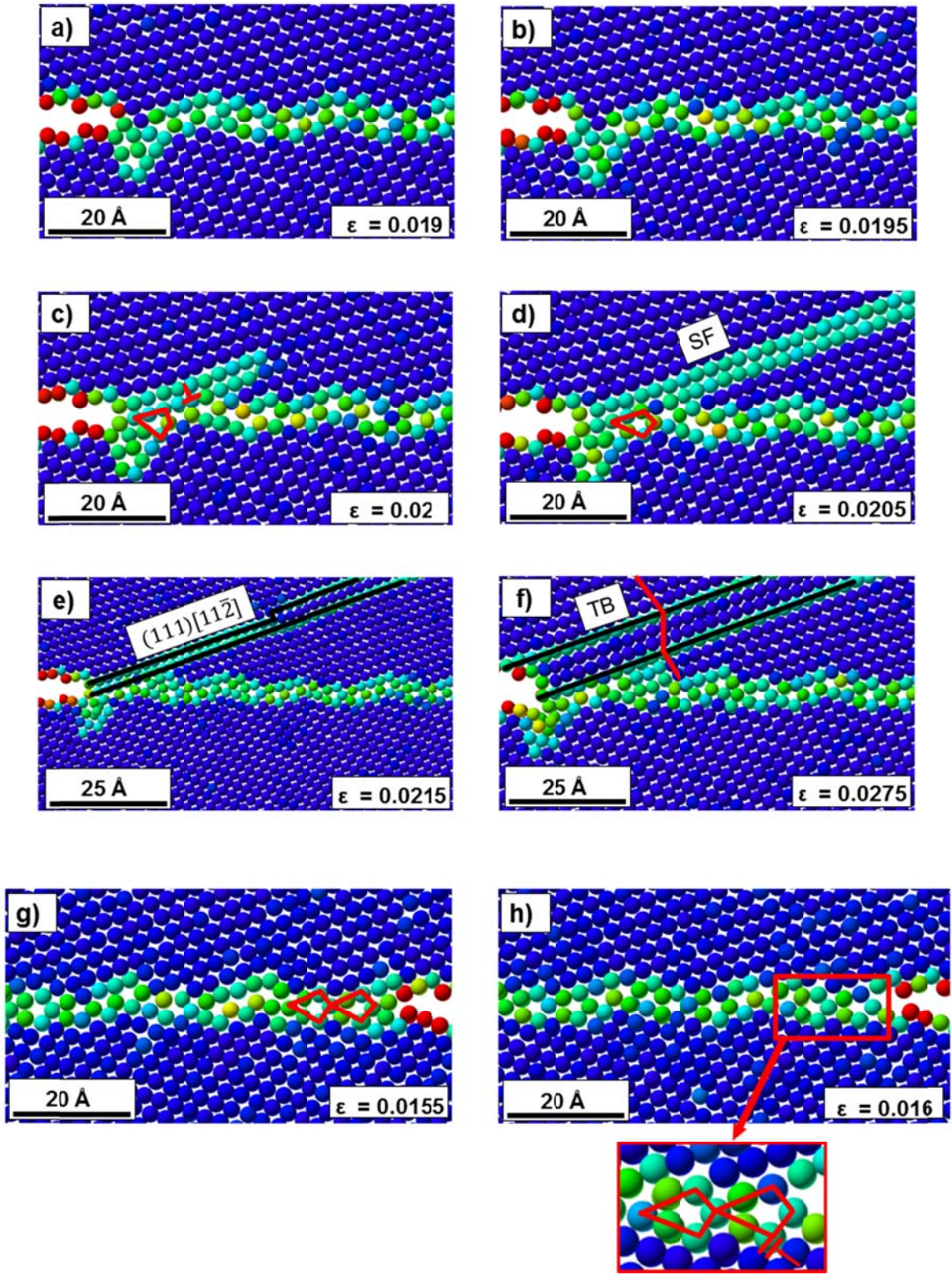

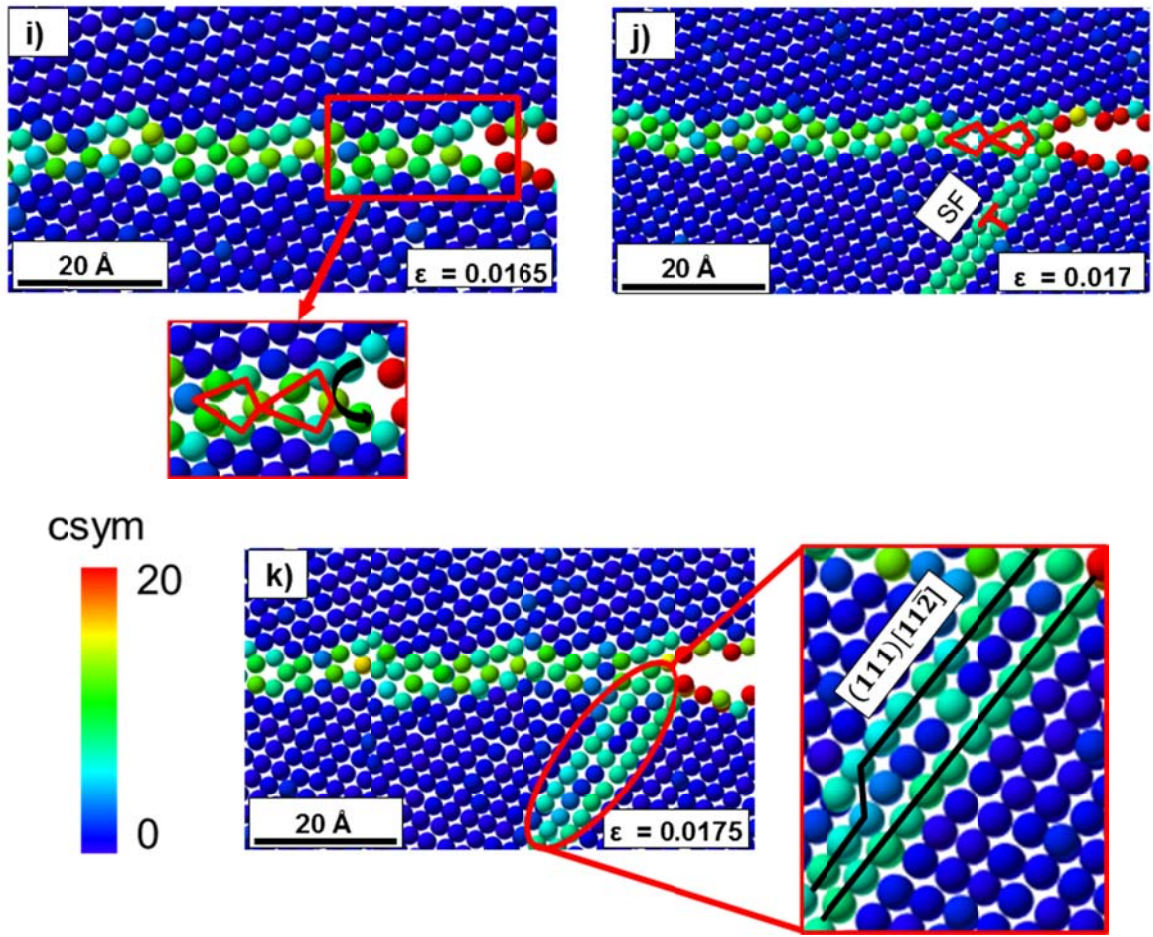

Figure 11. The plastic events across the Σ27 (552) GB interface with atoms are identified using centrosymmetry. (a-f) Show a nucleation of twin followed by growth of the crack tip along the +X direction, and (g-k) illustrate the twin nucleation and growth of the crack tip along the −X direction. The insets magnify the area of interest which depicted partial dislocation nucleation event followed by the twin formation.

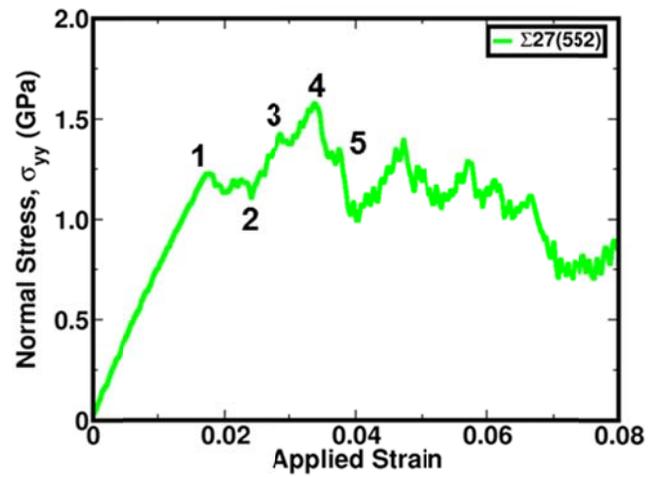

Figure 12. Mechanical response and highlights of key events (1-5) taking place during the deformation of Σ27 (552) θ=31.58° GB interface.

Table 1. Grain boundary CSL, SUM notation along with the GB periodic length, GB energy, the free volume and the maximum interface normal strength for each <100> STGB used in this work. The Σ97 (940) θ=47.9° GB has the highest initial free volume, periodic length, GB energy and displayed least interface normal strength among <100> STGBs examined here.

| Misorientation angle (θ) | CSL designation (Σ) | SUM notation | GB periodic length (Å) | GB energy (mJ/m$^2$) | Free volume ($\Delta z^*$) in units of $a_0$ | Max. interface normal strength $\sigma_{yy}^{max}$ (GPa) |
|---|---|---|---|---|---|---|
| 22.6° | Σ13 (510) | \|CDD\| | 20.65 | 542 | 0.1754 | 2.32 |
| 36.9° | Σ5 (310) | \|C\| | 12.80 | 551 | 0.1670 | 2.67 |
| 47.9° | Σ97 (940) | \|B''B''C'. B''B''C'\| | 39.08 | 596 | 0.2514 | 2.14 |
| 53.1° | Σ5 (210) | \|B'.B'\| | 9.05 | 565 | 0.0945 | 2.86 |
| 67.4° | Σ13 (320) | \|AB'.AB'\| | 14.60 | 481 | 0.2250 | 2.56 |

Table 2. Grain boundary CSL description, SU notation, the GB periodic length, GB energy, the free volume and the maximum interface normal strength associated with each <110> STGB used in this work. The Σ27 (552) GB has the highest initial free volume, periodic length, GB energy and lowest interface strength among <110> STGBs examined here.

| Misorientation angle (θ) | CSL designation (Σ) | SUM notation | GB periodic length (Å) | GB energy (mJ/m$^2$) | Free volume ($\Delta z^*$) in units of $a_0$ | Max. interface normal strength $\sigma_{yy}^{max}$ (GPa) |
|---|---|---|---|---|---|---|
| 50.48° | Σ11(113) | \|C.C\| | 18.91 | 151 | 0.1811 | 2.70 |
| 59° | Σ33 (225) | \|C.CDC.C\| | 32.75 | 326 | 0.2060 | 2.46 |
| 109.48° | Σ3 (111) | \|D.D\| | 9.87 | 73 | 0.0005 | 2.40 |
| 129.52° | Σ11 (332) | \|DE.DE\| | 26.75 | 390 | 0.1724 | 1.67 |
| 141.06° | Σ9 (221) | \|E.E\| | 17.10 | 437 | 0.1500 | 1.62 |
| 148.42° | Σ27 (552) | \|AEE.AEE\| | 41.90 | 474 | 0.2650 | 1.58 |
| 159.96° | Σ33 (441) | \|ED.ED\| | 32.75 | 428 | 0.1610 | 1.65 |

Table 3. Crack propagation figures of merit along the +X and –X directions for <100> STGB interfaces.

| CSL designation (Σ) | Maximum crack growth period along the -X direction (ps) | Maximum crack growth along the -X direction (Å) | Maximum crack growth velocity along the -X direction (m/s) | Maximum crack growth period along the +X direction (ps) | Maximum crack growth along the +X direction (Å) | Maximum crack growth velocity along the +X direction (m/s) |
|---|---|---|---|---|---|---|
| Σ13 (510) | 425 – 460 | 68 | 100 | 450 – 540 | 109 | 55 |
| Σ5 (310) | 340 – 500 | 32 | 19 | 380 – 400 | 21 | 50 |
| Σ97 (940) | 290 – 350 | 31 | 45 | 300 – 420 | 99 | 75 |
| Σ13 (320) | 360 – 500 | 24 | 8.5 | 300 – 410 | 34 | 27 |
| Σ5 (210) | 375 – 410 | 37 | 57 | 380 – 450 | 51 | 50 |

Table 4. Crack propagation figures of merit along the +X and –X directions for <110> STGB interfaces.

| CSL designation (Σ) | Maximum crack growth period along the -X direction (ps) | Maximum crack growth along the -X direction (Å) | Maximum crack growth velocity along the -X direction (m/s) | Maximum crack growth period along the +X direction (ps) | Maximum crack growth along the +X direction (Å) | Maximum crack growth velocity along the +X direction (m/s) |
|---|---|---|---|---|---|---|
| Σ11 (113) | 370 – 500 | 76 | 48 | 375 – 600 | 6 | 12 |
| Σ33 (225) | 570 – 615 | 86 | 85 | 320 – 510 | 62 | 22 |
| Σ3 (111) | 400 – 650 | 33 | 10 | 300 – 325 | 6 | 20 |
| Σ11 (332) | 360 – 500 | 17 | 10 | 250 – 290 | 4 | 10 |
| Σ9 (221) | NA | 3 | - | 300 – 380 | 12 | 12 |
| Σ27 (552) | 570 – 600 | 13 | 24 | 610 – 650 | 48 | 50 |
| Σ33 (441) | NA | 2 | - | NA | 0 | - |

Table 5. Summary of deformation modes observed in various <110> STGBs.

| CSL designation (Σ) | Structural unit (SU) | Mode of deformation |
|---|---|---|
| Σ3 (111) | |D.D| | Twinning |
| Σ11 (332) | |DE.DE| | Twinning |
| Σ9 (221) | |E.E| | Twinning |
| Σ11 (113) | |C.C| | Partial dislocation |
| Σ27 (552) | |AEE.AEE| | Twinning |
| Σ33 (441) | |ED.ED| | Twinning |
| Σ33 (225) | |C.CDC.C| | Partial dislocation |

Table 6. Summary of deformation modes observed in various <100> STGBs.

| CSL designation (Σ) | Structural unit (SU) | Mode of deformation |
|---|---|---|
| Σ5 (210) | |B'.B'| | Partial Dislocation |
| Σ5 (310) | |C| | Partial dislocation |
| Σ13 (510) | |CDD| | Partial dislocation |
| Σ13 (320) | |AB'.AB'| | Partial dislocation |
| Σ97 (940) | |B''B''C'. B''B''C'| | Partial dislocation |